\documentstyle[sprocl,amssymb]{article}

\input{epsf}
\newcommand{\CC}{{\Bbb C}}

\newcommand{\ZZ}{{\Bbb Z}}

\newcommand{\RR}{{\Bbb R}}
\newcommand{\QQ}{{\Bbb Q}}

\newcommand{\h}{\mbox{\small $\#$}}

\newcommand{\cf}{{\mathbf 1}}
\newcommand{\dg}{\dagger}

\newcommand{\ds}{\mathop{\mbox{\large $\amalg$}}}

\newtheorem{defin}{Definition}

\newcommand{\be}{\begin{equation}}
\newcommand{\ee}{\end{equation}}

\begin{document}

\title{\centerline{MULTI-COMPONENT MODEL SETS}
       \centerline{AND INVARIANT DENSITIES}}

\author{\sc Michael Baake {}\footnote{Heisenberg-fellow}}

\address{Institut f\"ur Theoretische Physik, Universit\"at T\"ubingen, \\
Auf der Morgenstelle 14, D-72076 T\"ubingen, Germany}

\author{\sc Robert V.~Moody}

\address{Department of Mathematical Sciences, \\ 
University of Alberta, Edmonton, Canada T6G 2G1}

\maketitle \abstracts{Model sets (also called cut and project sets) are
generalizations of lattices, and multi-component model sets are
generalizations of lattices with colourings. In this paper, we study
self-similarities of multi-component model sets. The main point may be simply
summarized:  whenever there is a self-similarity, there are also 
naturally related density functions. 
As in the case of
ordinary model sets \cite{BM3}, we show that invariant densities
exist and that they produce absolutely continuous invariant measures in
internal space, these features now appearing in matrix form.  
We establish a close connection between the theory of invariant densities
and the spectral theory of matrix continuous refinement operators.}

\section{Introduction}
Mathematical quasicrystals are tilings or Delone sets obtained from
the cut and project set or any other construction that ensures a
pure point diffraction spectrum. In many examples, they are one-component
sets, in the sense that their vertices (respectively their points)
 form just one
translation class with respect to the limit translation module \cite{BS} 
-- one well-known exception being the rhombic Penrose tiling whose
vertices fall into four different classes.

In view of the fact that practically all known physical quasicrystals
are multi-atomic alloys, a systematic treatment of multi-component
Delone sets is necessary and should give extra insight into the
phenomena possible for non-periodically ordered structures, such
as symmetries, inflation structure, etc. 
To make some progress here, one can use a generalized
set-up of internal space~\cite{Martin2}, and/or study the mutual
dependence of the components under inflations.        

Here, it is our aim to extend
recent work on the existence of inflation invariant measures on
such sets from the one-component to the multi-component case.
After introducing multi-component model sets and the notion
of affine self-similarities for them, we show that the set of all
affine self-similarities, based on a given linear inflation similarity, leads
directly to the concept of an invariant density on the points
of the model set. Determining this invariant
density necessitates moving the picture to the internal side where
it becomes a problem of finding fixed points of a matrix
refinement operator. We illustrate the nature of the solutions by
examining the situation in a pair of examples based on the
vertex set of a rhombic Penrose tiling, a 4-component model set.

\section{Multi-component model sets}
We begin with the notion of a normal cut-and-project scheme. By definition,
this consists of a collection of spaces and mappings: 

\be \label{cutandproject}
  \begin{array}{ccccc}
   \RR^m & \stackrel{\pi^{}_1}{\longleftarrow} & \RR^m \times \RR^n &
           \stackrel{\pi^{}_2}{\longrightarrow} & \RR^n  \\
    & & \cup & & \\ & & \tilde{L} & & \end{array}
\ee
where $\RR^m$ and $\RR^n$ are two real spaces, $\pi^{}_1$ and $\pi^{}_2$
are the projection mappings onto them, and 
$\tilde{L} \subset \RR^m \times \RR^n$
is a lattice. We assume that $\pi^{}_1|^{}_{\tilde{L}}$ is injective and
that $\pi^{}_2(\tilde{L})$ is dense in $\RR^n$. We call $\RR^m$
(resp.\ $\RR^n$) the physical (resp.\ internal) space. We will assume that
$\RR^m$ and $\RR^n$ are equipped with Euclidean metrics and that
$\RR^m \times \RR^n$ is the orthogonal sum of the two spaces. For $x$
lying in any of these spaces, $|x|$ denotes its Euclidean length.

Let $L := \pi^{}_1(\tilde{L})$ and let
\be \label{star}
     (.)^* \, : \quad L \; \longrightarrow \; \RR^n
\ee
be the mapping $\pi^{}_2 \circ (\pi^{}_1|^{}_{\tilde{L}})^{-1}$.
This mapping extends naturally to a mapping on the rational span 
$\QQ L$ of $L$, also denoted by $(.)^*$.

Let us now assume that $\tilde{L}$ is a sublattice of another lattice 
$\tilde{M}$, also lying in $\RR^{m+n}$. Necessarily, 
${\rm rank} _{\ZZ} ({\tilde M}) = m+n = 
{\rm rank} _{\ZZ}({\tilde L})$, so $[{\tilde M}:{\tilde L}]$ is finite. 
It is immediate that the pair $(\RR ^{m+n}, {\tilde M})$ gives 
rise to another cut and project scheme
and a new $\ZZ$-module $M:= \pi_1({\tilde M})$ of finite rank equal 
to $m+n$. Clearly, $M$ lies in the rational span of $L$ and 
$[M:L] = [{\tilde M}:{\tilde L}]$.

We now choose $r$ distinct {}\footnote{Distinctness is by no means
necessary, and it might actually be useful for some applications
to drop this assumption which makes no difference to the later
arguments.}  cosets $L^{(i)} := z^i+L$, $r \ge 1$, of $L$ in $M$ 
and $r$ windows $\Omega^{(1)}, \dots, \Omega^{(r)}$  inside the internal space
$\RR^n$. Throughout this paper, we will make the following assumptions 
on all windows $\Omega$,
in particular on the $r$ windows that we have just chosen:
\begin{itemize}
\item[\bf W1] $\Omega \subset \RR^n$ is compact;
\item[\bf W2] $\Omega \;=\; \overline{\mbox{int}(\Omega)}$;
\item[\bf W3] The boundary of $\Omega$ has Lebesgue measure 0.
\end{itemize}
With this data, we now define $r$ subsets of $\RR^m$ by:

\be
     \Lambda^{(i)} \; := \; \Lambda^{(i)}(\Omega^{(i)}) \; := \; 
                     \{ x \in L^{(i)} \mid  x^* \in \Omega^{(i)} \, \} \, .
\ee
Since we are assuming that the cosets $L^{(i)}$ are distinct, the 
$\Lambda^{(i)}$ are mutually disjoint. 
There are two ways in which we can view the collection of sets $\Lambda^{(i)}$.
The {\em flat} picture is
\be \label{phsyical}
    \Lambda \; := \; \bigcup_{i=1}^{r} \Lambda^{(i)} 
            \; \subset \; \RR^m
\ee
in which we view the sets simultaneously in $\RR^m$, as one would 
ultimately
need for atomic models. The second, and mathematically more useful, way
is the {\em fibred} picture in which we embed
the sets $\Lambda^{(i)}$ into $r$ copies of $\RR^m$:
\be \label{mcms}
    \Lambda^\dg \; :=\; \ds_{i=1}^{r}\Lambda^{(i)} 
                \; \subset \; \ds_{i=1}^{r} \RR^m,
\ee
We call this fibred picture a {\em multi-component} model set.

Let us pause to mention some of the nice properties of such point 
sets in the flat picture.
\begin{itemize}
\item[\bf M1] $\Lambda$ is a Delone set, i.e.\ $\Lambda$ is uniformly
  discrete and relatively dense.
\item[\bf M2] $\Lambda$ is a {\em Meyer} set: in addition to {\bf M1}
 there is a finite set $F$ so that 
 $\Lambda - \Lambda \subset \Lambda - F$.
 Equivalently, $\Lambda - \Lambda$ is also a Delone set.
\item[\bf M3] $\Lambda$ has a well-defined density $d$, i.e.\
\be \label{1.4}
     d \; = \; \lim_{s\rightarrow\infty}
        \frac{\h \Lambda_s}{\mbox{vol}(B_s(0))}
       \; = \; \lim_{s\rightarrow\infty}
        \frac{\h \Lambda_s}{c_m s^m}
\ee
exists, where $B_s(0) := \{ x \in \RR^m \mid |x| \leq s \}$, 
~$\Lambda_s:= \Lambda \cap B_s(0)$,
 and
$c_m \; := \; \frac{\pi^{m/2}}{\Gamma({m\over 2} + 1)}$
is the volume of the unit sphere in $\RR^m$.
\item[\bf M4] $\Lambda$ diffracts, i.e.\ it has a well-defined
Bragg spectrum.
\end{itemize}

These facts can be demonstrated by slightly rearranging the setup of
the cut and project formalism and then appealing directly to the papers
of Schlottmann \cite{Martin2} and Hof \cite{Hof} where they are proved. 
Set $G:=\tilde{M}/\tilde{L}$
and let $\alpha:\tilde{M} \rightarrow G$ be the canonical quotient map. 
We create a generalized cut and project scheme:
\be \label{gencutandproject}
  \begin{array}{ccccc}
   \RR^m & {\longleftarrow} & \RR^m \times (\RR^n \times G) &
          {\longrightarrow} & \RR^n \times G \\
    & & \cup & & \\ & & {\tilde N} & & \end{array}
\ee
where ${\tilde N}:= \{(x,\alpha(x)) \mid x\in {\tilde M}\} 
           \subset (\RR^m \times \RR^n) \times G
      = \RR^m \times (\RR^n \times G)$ with 
\be
    \pi_1(x) \; \leftarrow \; (x,\alpha(x)) 
                \; \rightarrow \; (\pi_2(x), \alpha(x)). 
\ee
Then, using the window 
\be
  \tilde{\Omega} \; = \; \bigcup_{j=1}^r \left(\Omega^{(j)} \times 
                         \alpha(z^j)\right)
\ee
we recover $\Lambda$ directly from (\ref{gencutandproject}).

The limit in (\ref{1.4}) is easily seen to be independent of the choice 
of origin for the Euclidean space; indeed, it even exists
{\em uniformly} for sets: for any subset $S$ of 
$\Omega$ with boundary of measure $0$, the relative frequency
of the points of $(\Lambda_s)^*$ falling into $S$, as $s \to \infty$, is 
${\rm vol}(S)/{\rm vol}(\Omega)$, and the convergence is uniform
with respect to translations of the set $S$.
See also Refs.~12, 13, and 11.

Multicomponent model sets are a natural and somewhat parallel 
algebraic counterpart
of tiling systems with various types of tiles or vertices. 
{}For example, in the case of the rhombic
Penrose tilings (see example below), it is well known \cite{BS}
that there are four classes of vertices corresponding to the four 
different cosets of the limit translation module (LTM), the class 
of the LTM itself not being present. These four classes can be 
viewed as algebraic in origin,
ultimately deriving from the structure of the prime ideal over 
$5$ in the ring of cyclotomic integers for the fifth roots of unity, 
see also Ref.~1.

As in the case of ordinary model sets, the generalization out of the
domain of ordinary lattices entails the loss of translational symmetry.
What often emerges instead, as one sees in many tiling models, is a very 
rich structure self-similarity. Our aim here is to begin to derive the 
analytical consequences of these new forms of symmetry.

\section {Self-similarities}

\begin{defin} A {\em self-similarity} of the fibred model set $\Lambda^\dg$ of
(\ref{mcms}) is an
$r\times r$ matrix $t = t_{Q,v}$ of maps 
\be \label{mcselfsim}
     t^{ji}\; : \; \Lambda^{(i)} \longrightarrow \Lambda^{(j)}
\ee
of the form 
\be
     t^{ji} \; : \; x \mapsto Qx+v^{ji},
\ee
where $Q$ is a fixed linear similarity map (i.e.\ a rotation followed by 
an inflation with a factor of $q$) stabilizing both $L$ and $M$, and the
$v^{ji}$ lie in $\RR^m$. 
\end{defin}
We call $Q$ the {\em linear part} of $t$ and the constant $q$ the 
{\em inflation factor} of $t$. 
Notice that this concept of self-similarity is not necessarily compatible
with any sensible mapping, on $\RR ^m$, of $\Lambda$ into itself, unless
the $v^{ji}$ are independent of $i,j$. The latter situation happens,
of course, in the one-component case \cite{BM3}.

Note that $Q$ naturally gives rise to an automorphism $\tilde{Q}$ of
the lattice $\tilde{M}$, i.e.\ an element of ${\rm GL}_{\ZZ}(\tilde{M})$,
and a linear mapping $Q^*$ of $\RR^n$ that maps
$\Omega$ into itself. From the arithmetic nature of $\tilde{Q}$
we deduce that the eigenvalues of $Q$ and $Q^*$ are algebraic
integers and from the compactness of $\Omega$ that 
$Q^*$ is contractive. Furthermore, one can deduce \cite{BM} that $Q^*$ is
diagonalizable from the corresponding property of $Q$. Strictly
speaking, we can only deduce that the eigenvalues of $Q^*$ do not exceed
$1$ in absolute value, but {\em we will always assume} that in fact
they are less than one in absolute value. Such linear transformations
induce so-called {\em hyperbolic transformations} on the associated torus
$\RR^{m+n}/{\tilde M}$. Since $Q$ also stabilizes $L$, ${\tilde Q}$
stabilizes ${\tilde L}$ and hence induces an automorphism of the group
$G = {\tilde M}/{\tilde L}$.
In the sequel, we will normally denote the contraction 
$Q^*$ by $A$ to match other sources.

To proceed, fix $Q$, once and for all, and look at the set 
${\cal T} = {\cal T}_Q$  of all self-similarities $t_v = t_{Q,v}$ 
of $\Lambda^\dg$ which have linear part $Q$. We define
\be \label{jiwindows}
   \Omega^{ji}:= \{u \in \RR^n \mid Q^*\Omega^{(i)} + u \subset \Omega^{(j)}\}.
\ee
We assume each $\Omega^{ji}$ is non-empty and  satisfies the 
window conditions [{\bf W}]
above. (In fact, it is not crucial that the second window condition 
be fulfilled, as we will explain below, and we will use this flexibility
in our examples. However, it
makes the exposition a little less cluttered to assume it).

We define 
\be 
\begin{array}{ccc}
L^{ji}  & := & L + (z^j - Q z^i) \\
T^{ji}  & := & \{ y\in L^{ji} \mid y^* \in \Omega^{ji} \}.
\end{array}
\ee
It is straightforward to see that for any matrix $r\times r$ matrix 
of elements of $\RR^m$ we have
\be
        t := t_v \in {\cal T}_Q ~ \; \Leftrightarrow \; ~ 
             v^{ji} \in T^{ji}\, , ~\; \mbox{{\rm for all}}~j,i.
\ee

An {\em invariant density} on $\Lambda^\dg$ is, by defintion, a set of 
non-negative functions 
$p^{\,j}:L^{(j)} \rightarrow \RR_{\ge 0}$, $j=1, \dots,r$, satisfying the 
following properties:
\begin{itemize}
\item[{\bf ID1}] The function $p^{\,j}$ vanishes off $\Lambda^{(j)}$;
\item[{\bf ID2}] The functions observe the equations
\be \label{invden2}
  p^{\,j}(x) \; = \; |\det (Q)| \cdot \lim_{s \to \infty} \sum_{i=1}^r 
                 \frac{\nu^{ji}}{\h T_s^{ji}}
                 \sum_{v\in T_s^{ji}} p^i(Q^{-1}(x-v)) \, ;
\ee
\item[{\bf ID3}] The following limits exist:
\be \label{invden3}
   w^j \; = \; \lim_{s \to \infty}
       \frac{\mbox{vol}\,(\Omega^{(j)})}{\h\, \Lambda_s^{(j)}} 
       \sum_{x \in \Lambda_s^{(j)}}p^{\,j}(x) \, ,
\ee
\end{itemize}
where $\nu = (\nu^{ji})$ is a non-zero matrix of non-negative numbers
and $w = (w^1, \dots, w^r)^t \neq 0$ is a vector (resp.\ an $r\times 1$ matrix)
of non-negative numbers. We will see that in order to solve 
for  such invariant densities, it is necessary
that the matrix $\nu$ and the vector $w$ satisfy:
\begin{itemize}
\item[{\bf PF1}] The spectral radius of $\nu$ is $1$, this is an eigenvalue
  of $\nu$, and  $w$ is the corresponding right Perron-Frobenius eigenvector.
\end{itemize}
The matrix $\nu$ encodes the degree of freedom to weight the points of one
coset relative to another. In view of this interpretation, 
it would be reasonable to choose $\nu$ to be a Markov matrix. 
In any case, the irreducibility of $\nu$, in the sense
of non-negative matrices, would be sufficient (though by
no means neccessary) to guarantee the additional property
\begin{itemize}
\item[{\bf PF2}] $1$ is a simple eigenvalue of $\nu$.
\end{itemize}
The right eigenvector $w$ (which has 
non-negative entries) may be normalized statistically,
i.e.\ $\sum_{j=1}^r w^j =1$, whereupon it is uniquely determined. 
Note that such a normalization takes
the relative densities of the points per unit volume into account,
while one could also work with a normalization per point. 

In effect, the main condition {\bf ID2} says that, apart from a
rescaling factor $|\det(Q)|$, the density $p^{\,j}(x)$ at a point
$x$ of $\Lambda^{(j)}$ is the averaged value of the densities
of the points $y$ which are mapped to $x$ by an element of
${\cal T}_Q$, after these averages have been weighted by the
non-negative matrix $\nu$. {\bf ID1} avoids the necessity of
determing which of the possible preimages $Q^{-1}(x-v)$ actually
lie in $\Lambda^{(i)}$ in (\ref{invden2}), and {\bf ID3} is a normalization
condition.

\section{Solving for the invariant densities}

We can solve for the invariant densities in the same manner as in Ref.~2.
We assume the existence of a set of {\em continuous} functions 
$f^j : \Omega^{(j)} \rightarrow
\RR_{\ge 0}$ such that $\mbox{supp}(f^j) \subset \Omega^{(j)}$ and
$f^j(x^*) = p^{\,j}(x)$ for all $x\in \Lambda^{(j)}$. We rewrite the equations
[{\bf ID}] on the window side. The uniform density of projection on the
window side allows us to use Weyl's theorem \cite{Kuipers,Weyl} to replace
the averaged sums by integrals.  To avoid undue proliferation
of symbols, we revert again to using $x$ as the variable, now living
in $\RR^n$. With $A:=Q^*$, this leads to 
\be \label{basicEqns}
\begin{array}{rcl}
f^j(x) &=& |\det(Q)| \,\sum_{i=1}^r \frac{\nu^{ji}}
                             {\mbox{vol}(\Omega^{ji})}
      \int_{\Omega^{ji}} f^i(A^{-1}(x-v))  dv \, ; \\
          ~ & ~ &~\\
      w^j &=& \int_{\Omega^{(j)}}f^j(x) dx \, .
\end{array}
\ee
Setting $X_S := \frac{\cf_S}{{\rm vol}(S)}$ for any measurable subset 
$S$ of $\RR^n$, we can rewrite the first equation in the form
\be \label{conv}
  f^j(x) \;=\; |\det (Q)|\,\sum_{i=1}^r \nu^{ji}
         \int_{\RR^n} X_{\Omega^{ji}}(x-y) \, f^i(A^{-1}y) dy  .\\
\ee
Defining $Y$ to be the matrix of functions $(\nu^{ji}X_{\Omega^{ji}})$
and $f = (f^1, \dots, f^r)^t$, we can write this more succinctly as
\be
f(x) = |\det(Q)| \, \int_{\RR^n} Y(x-y) f(A^{-1}y) dy
\ee
which expresses it in the form of a `matrix convolution'. 

We can solve this system of integral equations by taking 
Fourier transforms, getting (with $A^t$ denoting the transpose of $A$)
the matrix equation:
\be \label{matFtEqn}
     \hat{f}(k) \; = \; \hat{Y}(k) \cdot \hat{f}(A^t k).
\ee
This recursive formulation leads at once to the form of the desired 
solution (in Fourier space):
\be \label{fourierSol}
     \hat{f}(k) \; =\;  \prod_{\ell=0}^\infty 
                \hat{Y}((A^t)^{\ell})(k) \cdot \hat{f}(0).
\ee
{}From $k=0$ in (\ref{matFtEqn}), we get the consistency equation
\be
  \hat{f}(0) \, = \,  w \, = \, \hat{Y}(0) \cdot \hat{f}(0) 
             \, = \, \nu \cdot \hat{f}(0) \, ,
\ee
so $w$ is an eigenvector for the matrix $\nu = (\nu^{ji})$
with eigenvalue $1$. This explains the necessity of the conditions 
[{\bf PF1}].

The matrix linear operator $R$ on the Banach space $L_r^2(\Omega)$ 
of $r \times 1$ 
column matrices of $L^2$ functions with support of the $i$th
component in $\Omega^{(i)}$  defined by
\be \label{cntsRefOp}
R(F) \;=\; |\det(Q)| \int_{\RR^n}Y(x-y)F(A^{-1}y) dy
\ee
is an example of a matrix continuous refinement operator in the sense
of Jiang and Lee \cite{Jiang}. We are looking for a $1$-eigenfunction
of this operator. The situation regarding the spectrum of $R$ 
requires some care because there are two matrices involved: $A$ and $\nu$. 
Under the assumptions [{\bf PF1}] a solution $g \in C^\infty(\Omega)$ 
exists.
If, in addition, [{\bf PF2}] holds then there exists a unique
$C^\infty$ solution  $g$ (see Ref.~8, Theorem 3.1 and Remark 1 following it). 
In our particular case, it is not too hard 
to see that the infinite product
(\ref{fourierSol}) converges uniformly on compact sets to a 
vector of $C^\infty$-functions. 
Now, ${\hat g}$ also satisfies the defining equation 
(\ref{matFtEqn}) and ${\hat g}(0)$ is also a $1$-eigenfunction of $\nu$,  
hence, after rescaling,
can be taken to be $w$. Then it is straightforward to see that in 
fact ${\hat g}$ is our ${\hat f}$ of (\ref{fourierSol}). {}Finally, 
taking the inverse Fourier transform of (\ref{fourierSol}), we obtain
\be
   g(x) \;= \; f(x) \;= \; 
          \left[ \; \mathop{\mbox{\Huge $*$}}_{\ell=0}^{\infty}
                   \, (Y \circ A^{\ell}) \, \right] (x) \cdot w \, ,
\ee
with obvious meaning to the convolution as applied to matrices.
This is then the {\em unique} solution to the invariant
density problem of (\ref{basicEqns}). Defining
\be
   p^{\,j}(x) \; := \; f^j(x^*)\, , ~ j = 1, \dots, r \, ,
\ee
solves the invariant density problem on the multi-component model set 
$\Lambda$. The functions $f^j$ may be viewed as the Radon-Nikodym
derivatives of a set of $r$ absolutely continuous measures on
the windows $\Omega^{(j)}$.

\section{Example of Penrose tilings}

The simplest multicomponent model set of relevance to quasicrystal theory is
the set of vertex points of the above mentioned rhombic Penrose tiling. 
Recall the algebraic construction of such vertex sets \cite{BKSZ}.
Let $\xi := e^{2 \pi i /5}$ and 
let $\ZZ[\xi] := \{\sum_{j=0}^r m_j \xi^j \mid r \ge 0, m_j \in \ZZ \} 
\subset \CC$. Clearly, $\ZZ[\xi]$ is the ring of integers of the cyclotomic 
field of the fifth roots of
unity and is a rank $4$ $\ZZ$-module with basis $\{1, \xi, \xi^2, \xi^3 \}$. 
We can embed $\ZZ[\xi]$ into $\CC^2 \simeq \RR^4$ by the 
mapping $x \mapsto (x, x^*)$, where $*$ is the Galois
automorphism $\ZZ[\xi]$ defined by sending $\xi$ to $\xi^2$.  
There is a homomorphism ~ 
$\varrho : \ZZ[\xi] \longrightarrow \ZZ/5 \ZZ$ ~
defined by  $\varrho(\sum m_j \xi^j) = \sum m_j \quad \mbox{mod} \;5\,$,
which is suitable to define cosets.

Let $P$ be the pentagon which is the convex hull of 
$\{1, \xi, \dots \xi^4\}$ and define the windows
$\Omega^{(1)} := P$, $\Omega^{(4)}:= -P$, 
$\Omega^{(3)} := \tau P$,  $\Omega^{(2)} := -\tau P$. 
The notation is chosen to be
compatible with the fact that $\varrho(\tau) = 3 \; \mbox{mod} \;5$.
Define $L^{(i)} := \{ x \in \ZZ[\xi] \mid \varrho(x) = i \}$ and observe that 
$L^{(1)}, \dots , L^{(4)}$ are four different cosets of $L^{(0)}$ in $\ZZ[\xi]$. 
We define
\be \label{penrose}
    \Lambda^{(i)} := \{ x \in L^{(i)} \mid x^* \in \Omega^{(i)} \}.
\ee
Then, $\Lambda = \bigcup \Lambda^{(i)}$ is the set of vertices of a (singular) 
Penrose tiling. More precisely, $\Lambda$ is the union of 10 such
singular sets, deviating from a regular set only in points of density 0,
due to window condition {\bf [W3]}.
To obtain regular tilings, it is necessary to add a displacement
$\gamma \in \CC$ to each of the windows $\Omega^{(i)}$ in (\ref{penrose}), 
but ultimately this makes no
difference to the measure theoretical considerations that are involved in the 
invariant densities on the windows, 
so we choose to suppress this  additional complication here. 

We take, as our fixed linear self-similarity, the scaling maps 
$Q = \tau 1_\CC$, with 
corresponding contraction $A = Q^* = -\frac{1}{\tau} 1_\CC$ on the window 
side. The transition windows of (\ref{jiwindows}) become (with $\RR^2 = \CC$)
\be
       \Omega^{ji} := \{ u \in \CC \mid -\frac{1}{\tau}\Omega^{(i)} +u  
                                                          \subset \Omega^{(j)} \} \, .
\ee
Determining these is a simple geometrical problem the outcome of which
can be summarized in the following table
\be
\left[ \begin{array}{cccc}
(1/\tau^3)\, P & \{0\} & \emptyset & (1/\tau^2)\, P \\
-P & -(1/\tau^2)\, P & -(1/\tau)\, P & -(1/\tau^3 + 1/\tau)\, P\\
(1/\tau^3 + 1/\tau)\, P & (1/\tau)\, P & (1/\tau^2)\, P & P\\
-(1/\tau^2)\, P &\emptyset & \{0\} & -(1/\tau^3)\,P
\end{array} \right]
\ee
where $P = \Omega^{(1)}$.
The $ji$ entry of the table is the region $\Omega^{ji}$. 
We meet here a natural situation in which some of the 
windows are degenerate (the empty and singleton windows)
and do not satisfy the window hypotheses. Since we rely on
Weyl's theorem to connect the physical and internal sides of the
picture, and this is not applicable in such cases, 
we have to omit transition cases where
${\rm vol}(\Omega^{ji})=0$. Otherwise, the normalized
characteristic functions create ``ghosts'', i.e.\ $X_{\Omega^{ji}}$
is no longer a function but becomes a Dirac measure. We adopt the
convention of choosing $\nu^{ji}=0$ whenever that happens.

We are still free to choose the transition matrix $\nu$. For the 
purposes of
illustration, we have chosen two examples and included figures that show 
the shapes of the corresponding invariant densities on the windows.

\subsection{Example 1}
In this example, we define 
\be
    \nu \; = \;      
          \left(\frac{ {\rm vol}( \Omega^{ji}) }
                        {\sum_{k=1}^4 {\rm vol} (\Omega^{ki}) } \right)
          \; = \; \left( \begin{array}{cccc}
             (2-\tau)/4 & 0 & 0 & (\tau-1)/4 \\
             \tau/4 & 2-\tau & \tau-1 & (3-\tau)/4 \\
             (3-\tau)/4 & \tau-1 & 2-\tau & \tau/4 \\
             (\tau-1)/4 & 0 & 0 & (2-\tau)/4 \end{array} \right)
\ee
in which the contribution of points of type $i$ to the density of points of type 
$j$ is weighted by the areas of the corresponding transition windows. 
In this case, $\nu$ is a Markov matrix (i.e.\ $\nu^{ji} \geq 0$
and $\sum_{j=1}^4 \nu^{ji} = 1$), so the PF eigenvalue is $1$,
with corresponding eigenvector $(0,1,1,0)^t$. We get non-vanishing
densities only on the two windows $\Omega^{(2)}$ and $\Omega^{(3)}$,
a phenomenon caused by $\nu$ being a reducible matrix (although condition
{\bf [PF2]} still holds).


\vspace*{5mm}
\begin{figure}[ht]
\centerline{\epsfysize=90mm \epsfbox{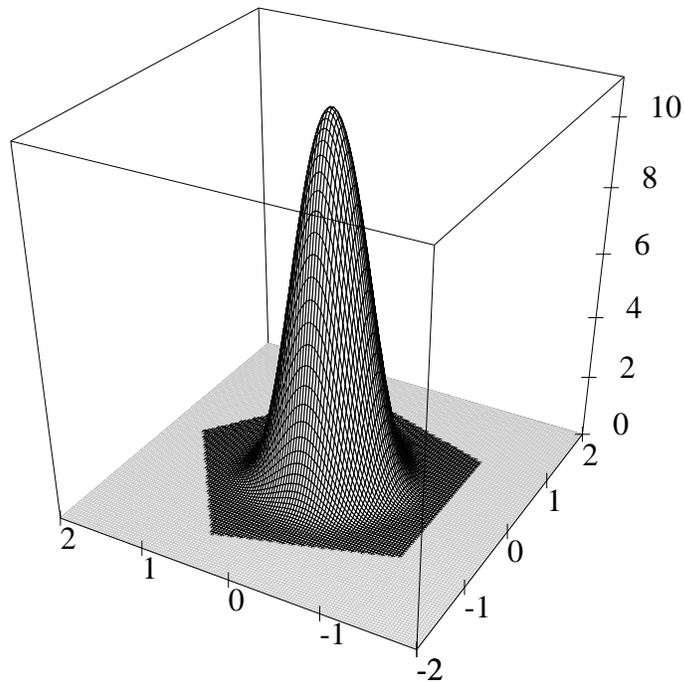}}
\caption{Invariant density on window $\Omega^{(2)}$ for Example 1.}
\end{figure}

\subsection{Example 2}
Here, we define
\be
    \nu \; = \;  {1\over 4} \left( \begin{array}{cccc}
                       2 & 0 & 0 & 2 \\
                       1 & 1 & 1 & 1 \\
                       1 & 1 & 1 & 1 \\
                       2 & 0 & 0 & 2 \end{array} \right) \, .
\ee
This $\nu$ is the transpose of a Markov matrix, whence the
PF eigenvalue is still $1$, now with right eigenvector $(1,1,1,1)^t$.
This means that, on the 4 windows, the integrated densities
must give equal values.


\vspace*{8mm}
\begin{figure}[ht]
\centerline{\epsfysize=55mm \epsfbox{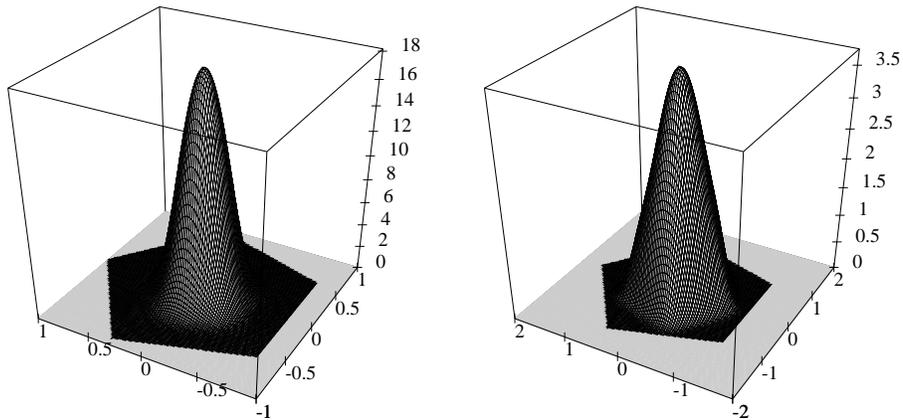}}
\caption{Invariant densities on windows $\Omega^{(1)}$ 
              and $\Omega^{(2)}$ for Example 2. Notice that the
two plots are scaled differently and that the invariant density
on the right is defined on the larger of the two windows.}
\end{figure}

\vspace*{3mm}
In the Figures, we show the invariant densities on only two of the windows, 
a large and a small pentagon. Because of the inherent symmetry in our 
examples, the densities on the other two windows differ only in orientation.
Though it is not necessary for $\nu$ to be Markov, the use of a Markov matrix, 
which preserves probabilities, seems appropriate for potential applications of
this picture to a probabilistic version with positive entropy density 
\cite{BM2,Dieter}. Further examples with different scaling factors,
in particular with $q=\tau^2$ where the invariant densities flatten out
at the top, will be given elsewhere \cite{BM}.

\section*{Acknowledgements}

It is our pleasure to thank Thomas D.\ Wilkinson for his 
considerable help preparing the Examples. This work was
supported by the German Science Foundation (DFG) and
by the Natural Sciences and Engineering Research Council
of Canada (NSERC).

\end{document}